\documentclass[
prd 
,showpacs ,amssymb,superscriptaddress,aps
]{revtex4}
\usepackage{graphicx}
\input{epsf}

\usepackage{amsmath,amssymb}
\usepackage{bm}
\usepackage{times}

\newcommand{\dalm}{\kern1pt\vbox{\hrule height 0.9pt\hbox{\vrule width 0.9pt
\hskip 2.5pt\vbox{\vskip 5.5pt}\hskip 3pt\vrule width 0.3pt}\hrule height 0.3pt}
\kern1pt}

\newcommand{\gsim}{\, \raisebox{-0.8ex}{$\stackrel{\textstyle >}{\sim}$ }}

\newcommand{\be}{\begin{eqnarray}}
\newcommand{\ee}{\end{eqnarray}}
\newcommand{\beq}{\begin{eqnarray}}
\newcommand{\eeq}{\end{eqnarray}}

\begin{document}



\title{Pulse profiles from a pulsar in scalar-tensor gravity}

\author{Hajime Sotani}
\email{sotani@yukawa.kyoto-u.ac.jp}
\affiliation{Division of Theoretical Astronomy, National Astronomical Observatory of Japan, 2-21-1 Osawa, Mitaka, Tokyo 181-8588, Japan}


\date{\today}

\begin{abstract}
The pulse profile from a neutron star in scalar-tensor theory of gravity is examined for several stellar models, where we assume the existence of the antipodal hot spots on the neutron star based on the polar cap model. Then, we find that the pulse profile from the scalarized neutron star in scalar-tensor gravity is almost the same as that in general relativity, i.e., without a scalar field, if the stellar compactness of the both stars is very similar. That is, the existence of the scalar field does not directly change the pulse profile from the neutron star, while the stellar compactness is crucial for determining the pulse shape even in the scalar-tensor gravity. Additionally, we find that the pulse shape from the scalarized neutron star is more or less similar to that from the neutron star with the same mass in general relativity, while the ratio of the minimum amplitude to the maximum amplitude in the pulse profile depend strongly on the coupling constant in scalar-tensor gravity, depending on the angle between the rotational and magnetic axes and the angle between the rotational axis and the direction to the observer. So, the direct observation of the pulse profile together with the additional observation of the stellar mass, one may extract the imprint of the gravitational theory in strong field regime.
\end{abstract}

%
\maketitle
\section{Introduction}
\label{sec:I}

Gravitational theory is one of the most fundamental theories in physics. Among various theories of gravity proposed so far, general relativity proposed by Einstein has been tested by many experiments and astronomical observations, which do not show any defects of the theory. However, most of the tests have been done in a weak gravitational field, such as the solar system. On the other hand, the verifications of general relativity in strong field regime inside/around the compact objects are very poor, i.e., the theory of gravity describing strong field regime may deviate from general relativity. If so, the phenomena associated with the compact objects might be different from the expectations in general relativity, which can become a probe of the gravitational theory in  strong field regime. The development of observational technologies would enable us to observe such phenomena, e.g., the gravitational waves from the compact objects, with surgical precision. Up to now, several possibilities for probing the theory of gravity have been suggested (e.g., \cite{Berti2015,DP2003,SK2004,Sotani2014,Sotani2014a}).

Neutron star, which is a byproduct via the core-collapsed supernova explosion, is a good candidate for testing the physics in the extreme conditions. The density inside the star significantly exceeds the standard nuclear density, the magnetic field inside/around the star can be very strong, and the field strength of gravity also becomes too strong \cite{shapiro-teukolsky}. Due to such a strong field of gravity, one could see a relativistic effect in the light bending radiated from the stellar surface. As a result, for example, the pulse profiles from the pulsar strongly depend on how strong the gravitational field is. So, the direct observations of the pulse profile may tell us the stellar compactness, which is the ratio of the stellar mass to the radius \cite{PFC1983,POC2014,Bogdanov2016}. 
This is one of the objectives in NICER project \cite{NICER}.
Moreover, the rotational effects in pulse profiles are also taken into account \cite{CLM05,PO2014}. Meanwhile, the properties of light bending should depend on the geometry of gravitational fields \cite{SM2017}, where we show the possibility that the pulse shapes for a static, spherically symmetric spacetime in vacuum outside the star, which is different from the Schwarzschild spacetime, could be different from those for the Schwarzschild spacetime. That is, one could adopt the observations of the pulse profile as a test of the theory of gravity.

Scalar-tensor theory of gravity is one of the most natural extensions of general relativity, where the field equations are constructed by the usual tensor field as well as a scalar field. Scalar-tensor gravity has been originally discussed in the 1950$-$1960s with the works by Jordan, Fierz, Brans, and Dicke \cite{J1949,F1956,J1959,B1961,D1962}. Then, Damour and Esposito-Far\`{e}se have extended with the different coupling between the physical and Einstein frames \cite{DE1993}, where a new phenomenon in the compact stars, the so-called spontaneous scalarization, has been revealed \cite{DE1993,H1998}. The spontaneous scalarization typically arises in the neutron star models with more compactness. When the spontaneous scalarization happens, the stellar models in the branch expected in general relativity are unstable. In this paper, we particularly focus on scalar-tensor gravity and examine whether the pulse profile can change due to the existence of a scalar field. In addition, we study how the pulse profiles from the pulsar expected in scalar-tensor gravity can deviate from those expected in general relativity.

In the next section, we briefly mention the neutron star models in scalar-tensor gravity. In Sec. \ref{sec:III}, we examine the photon trajectory and deflection angle in scalar-tensor gravity. Then, in Sec. \ref{sec:IV} we present the numerical results of the pulse profiles from a pulsar, which are compared with the expectation in general relativity. Finally, we conclude in Sec. \ref{sec:V}. In this paper, we adopt geometric units, $c=G_*=1$, where $c$ and $G_*$ denote the speed of light and the gravitational constant, respectively, and the metric signature is $(-,+,+,+)$.

\section{Neutron stars in Scalar-tensor gravity}
\label{sec:II}

In this paper, we consider a neutron star model in scalar-tensor gravity with one scalar field $\varphi$, where gravity is mediated by a usual metric tensor together with a massless long-range scalar field. The action describing scalar-tensor gravity is given by
\begin{equation}
  S=\frac{1}{16\pi G_*}\int\sqrt{-g_*}\left(R_* - 2g_*^{\mu\nu}\varphi_{,\mu}\varphi_{,\nu}\right)d^4x + S_{\rm m}\left[\Psi_{\rm m},{\cal A}^2g_{*\mu\nu}\right],  \label{eq:action}
\end{equation}
where $G_*$ is the bare gravitational coupling constant, $R_*$ is the scalar curvature constructed with the Einstein metric $g_{*\mu\nu}$, $g_*$ is the determinant of $g_{*\mu\nu}$, and $S_{\rm m}$ denotes the action for all matter fields expressed by $\Psi_{\rm m}$ collectively. We remark that this action is expressed in the Einstein frame, which is not a physical frame, because the field equations in this frame become much simpler than those in the physical frame. On the other hand, the experimental phenomena are explained in the physical frame, which is sometimes referred to as the Jordan frame. To clarify the frame of the quantities, we denote the quantities in the Einstein frame with asterisk, while those in the physical frame with tilde. The metric in the physical frame, $\tilde{g}_{\mu\nu}$, is associated with $g_{*\mu\nu}$ via 
\begin{equation}
  \tilde{g}_{\mu\nu} = {\cal A}^2(\varphi)g_{*\mu\nu}.  \label{eq:EJ}
\end{equation}

The field equations in the Einstein frame can be derived by varying the action given by Eq. (\ref{eq:action}) with respect to $g_{*\mu\nu}$ and $\varphi$,
\begin{gather}
   G_{*\mu\nu} = 8\pi G_* T_{*\mu\nu} + 2\left(\varphi_{,\mu}\varphi_{,\nu} - \frac{1}{2}g_{*\mu\nu}g^{\alpha\beta}_*\varphi_{,\alpha}\varphi_{,\beta}\right), \\
   \dalm_*\varphi = -4\pi G_* \alpha(\varphi)T_*.
\end{gather}
Here, $T_{*\mu\nu}$ denotes the energy-momentum tensor in the Einstein frame, which is related to the energy-momentum tensor in the physical frame $\tilde{T}_{\mu\mu}$ as
\begin{equation}
  T_*^{\mu\nu} \equiv \frac{2}{\sqrt{-g_*}}\frac{\partial S_m}{\partial g_{*\mu\nu}} = {\cal A}^6\tilde{T}^{\mu\nu}.
\end{equation}
Additionally, $T_*$ and $\alpha(\varphi)$ are 
\begin{gather}
  T_* \equiv T_{*\mu}^{\ \ \ \mu} = g_{*\mu\nu}T_*^{\mu\nu}, \\
  \alpha(\varphi) \equiv \frac{d\ln {\cal A}(\varphi)}{d\varphi}.
\end{gather}
We remark that the scalar-tensor gravity in the limit of $\alpha=0$ reduces to general relativity. The energy-momentum conservation should be satisfied in the physical frame, i.e., $\tilde{\nabla}_\nu\tilde{T}_{\mu}^{\ \ \nu}=0$, which leads to the equation in the Einstein frame such as
\begin{equation}
  \nabla_{*\nu} T_{*\mu}^{\ \ \ \nu} = \alpha(\varphi) T_*\nabla_{*\nu}\varphi.
\end{equation}

With respect to the conformal factor in Eq. (\ref{eq:EJ}), we adopt the same functional form as in Ref. \cite{DE1993}, such as
\begin{equation}
  {\cal A}(\varphi) = \exp\left(\frac{1}{2}\beta\varphi^2\right),
\end{equation}
where the coupling constant $\beta$ is a real number. With this conformal factor, $\alpha = \beta\varphi$, i.e., the scalar-tensor gravity reduces to general relativity when $\beta = 0$. The observational constraint on the value of $\beta$ is quite severe. That is, the observations of the binary system composed of a neutron star and a white dwarf suggest that $\beta$ should be in the range of $\beta \gsim -5$ \cite{Freire2012}. On the other hand, theoretical studies tell us that the spontaneous scalarization sets in for $\beta\lesssim -4.35$ for spherically symmetric neutron stars \cite{H1998}, and for $\beta\lesssim -3.9$ for extremely fast-rotating neutron stars \cite{DYSK2013}. Moreover, during the evolution of the binary neutron stars system, the different type of the scalarization may take place for large value of $\beta$ \cite{PBPL2014,TSB2015}, which is referred to as the dynamical scalarization. In any case, considering to the observational constraint on $\beta$, the allowed range of $\beta$ with which the scalarization can be observed may not be so large.

A neutron star model in scalar-tensor gravity can be constructed by integrating the modified Tolman-Oppenheimer-Volkoff equations together with the appropriate equation of state (EOS) \cite{SK2004,H1998}. In order to determine the radial distribution of the scalar field, one has to adopt an asymptotic value of the scalar field $\varphi_0$. In particular, we focus on the case with $\varphi_0=0$ in this paper. Then, the neutron star model becomes one parameter family, e.g., with respect to the central density $\tilde{\rho}_c$ or the Arnowitt-Deser-Misner (ADM) mass of the star, by fixing the EOS and the coupling parameter $\beta$. In this paper, we adopt two EOSs, i.e., the so-called FPS EOS \cite{FPS} and Shen EOS \cite{Shen}. FPS EOS is based on the Skyrme type effective interaction, while Shen EOS is based on the relativistic mean field approach. The expected maximum masses of neutron star in general relativity are $1.80M_\odot$ for FPS EOS and $2.21M_\odot$ for Shen EOS.

We remark that FPS EOS may be ruled out by the observations of $2M_\odot$ neutron star \cite{D2010,A2013}. On the other hand, the saturation parameters of Shen EOS are not compatible with those experimentally observed on the Earth, in particular the density-dependent symmetry energy, the so-called slope parameter \cite{LL2013,O2017}. Even so, the both EOSs we adopt in this study are an example of extreme cases for the soft and stiff EOSs. That is, the stiffness of most of the realistic EOSs are between those of FPS EOS and Shen EOS, which leads to the result that the mass and radius expected with most of the realistic EOSs are plotted between the mass-radius relations constructed with FPS EOS and Shen EOS. Then, in Sec. \ref{sec:IV} we will show that the pulse profile from the scalarized neutron star with a specific EOS (FPS EOS) is very similar to that with another EOS (Shen EOS) if the stellar compactness is similar to each other. To show this result, it is better to adopt the extreme EOSs even though it is not suitable from the astronomical observations and terrestrial experiments. With such a reason, we especially adopt FPS EOS and Shen EOS in this study.

\begin{figure*}
\begin{center}
\begin{tabular}{cc}
\includegraphics[scale=0.5]{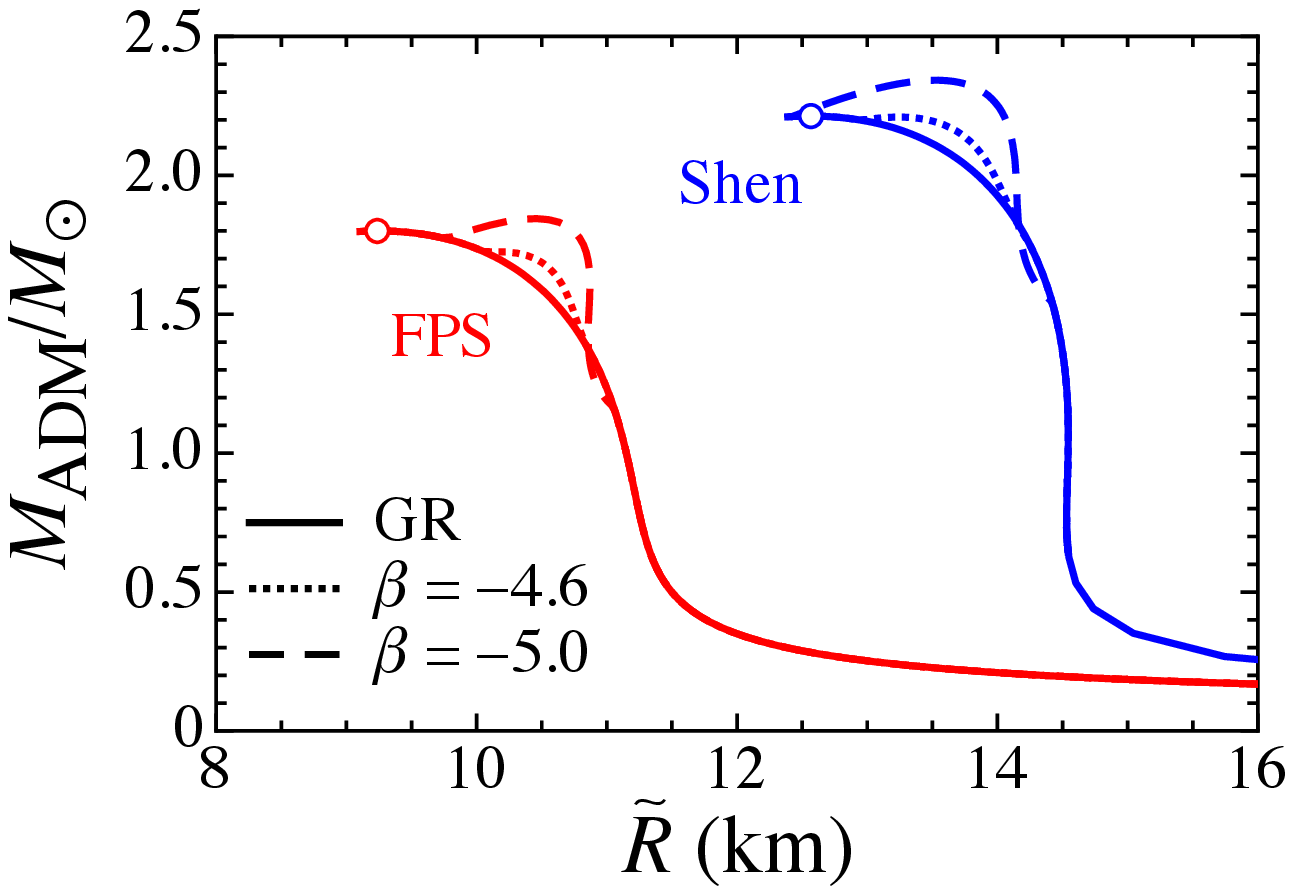} &
\includegraphics[scale=0.5]{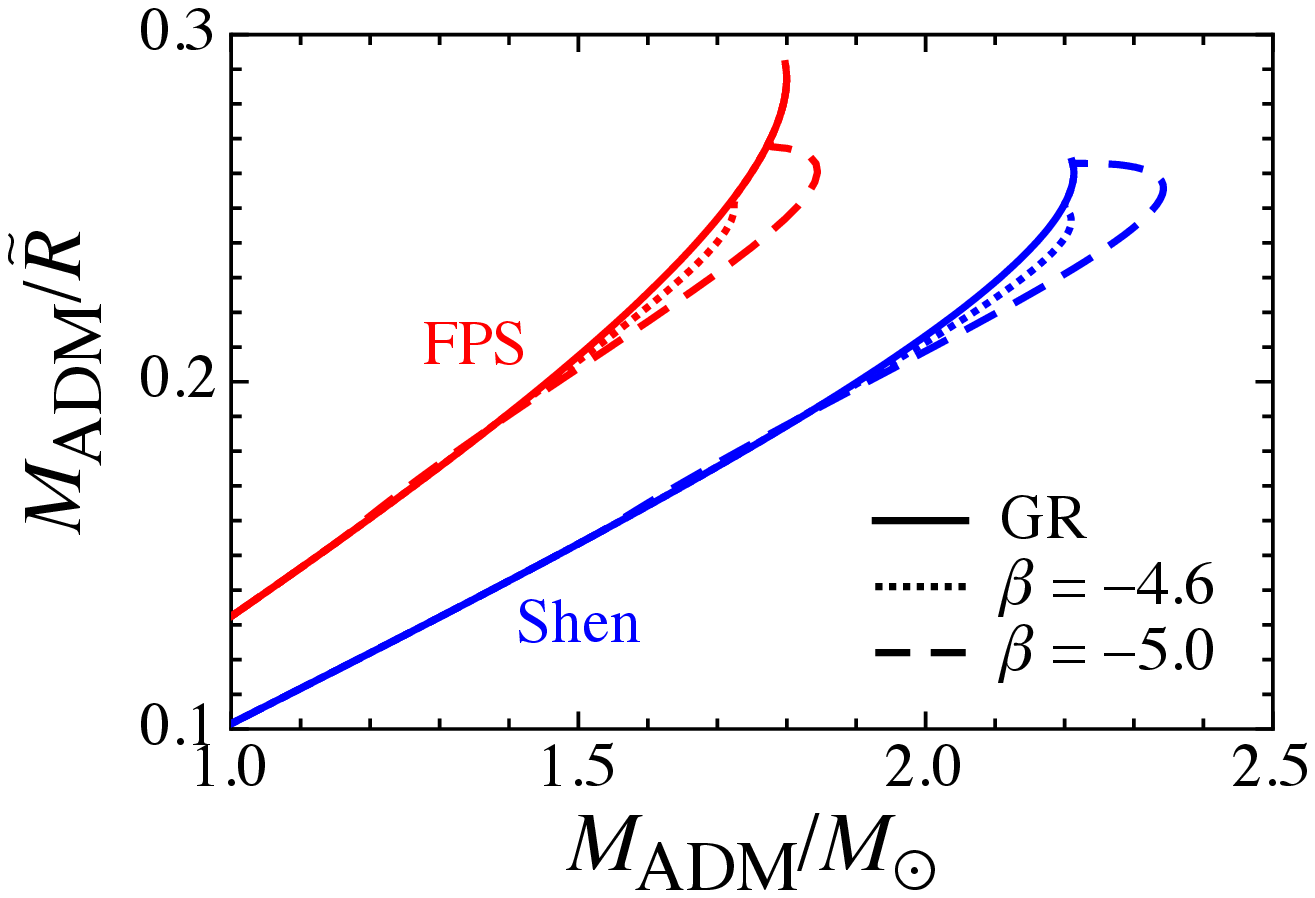} 
\end{tabular}
\end{center}
\caption{
Mass-radius relations of the neutron stars constructed with FPS and Shen EOSs are shown in the left panel. The right panel is the stellar compactness $M_{\rm ADM}/\tilde{R}$ as a function of the ADM mass. The solid lines denote the results in general relativity, while the dotted and dashed lines denote the results in scalar-tensor gravity with $\beta=-4.6$ and $-5.0$. The open circles correspond to the stellar models with the maximum mass in general relativity.
}
\label{fig:MR}
\end{figure*}

In Fig. \ref{fig:MR}, we show the relation between the ADM mass ($M_{\rm ADM}$) and the stellar radius ($\tilde{R}$) in the left panel, and the relation between the stellar compactness ($M_{\rm ADM}/\tilde{R}$) and the ADM mass in the right panel with respect to the neutron star models constructed with FPS and Shen EOSs. In this figure, the solid lines denote the results in general relativity, while the dotted and dashed lines respectively denote the results in scalar-tensor gravity with $\beta=-4.6$ and $-5.0$. We remark that the stellar compactness is an important property for discussing the light bending, because the relativistic effect becomes more important as the stellar compactness increases. In addition, the open circles in the left panel of the same figure correspond to the stellar models with the maximum mass in general relativity. The stellar models left to these marks in the left panel become unstable against the radial perturbations.

\section{Photon trajectory and deflection angle}
\label{sec:III}

We consider the photo trajectory from a spherically symmetric neutron star in scalar-tensor gravity, according to Ref. \cite{SM2017}. The metric in the physical frame is generally written as
\begin{equation}
  d\tilde{s}^2 = -\tilde{A}(r)dt^2 + \tilde{B}(r)dr^2 + \tilde{C}(r)(d\theta^2 + \sin^2\theta d\psi^2). \label{eq:metric}
\end{equation}
In particular, we adopt the metric form in the Einstein frame as
\begin{equation}
  ds_*^2 = -e^{2\Phi(r)}dt^2 + e^{2\Lambda(r)}dr^2 + r^2(d\theta^2 + \sin^2\theta d\psi^2).
\end{equation}
Then, via Eq. (\ref{eq:EJ}), the metric functions in Eq. (\ref{eq:metric}) become
\begin{align}
  \tilde{A}(r) &= \exp\left(2\Phi+\beta\varphi^2\right), \\
  \tilde{B}(r) &= \exp\left(2\Lambda + \beta\varphi^2\right), \\
  \tilde{C}(r) &= r^2 \exp\left(\beta\varphi^2\right).
\end{align}
Since the circumference radius $r_c$ is related to the metric function $\tilde{C}(r)$, i.e., $r_c^2=\tilde{C}(r)$, the stellar radius $\tilde{R}$ is given by $\tilde{R}^2=\tilde{C}(R)$. Hereafter, we express $\tilde{R}$ and $R$ as the stellar radius in the physical frame and the corresponding position in the radial coordinate of $r$, respectively. Owing to the condition of $\varphi_0=0$, the asymptotic behavior of the metric functions become
\begin{align}
  \tilde{A}(r) &= 1-\frac{2M_{\rm ADM}}{r} + {\cal O}\left(\frac{1}{r^2}\right), \\
  \tilde{B}(r) &= 1+\frac{2M_{\rm ADM}}{r} + {\cal O}\left(\frac{1}{r^2}\right), \\
  \tilde{C}(r) &= r^2 \left[1+{\cal O}\left(\frac{1}{r^2}\right)\right],
\end{align}
where $M_{\rm ADM}$ denotes the ADM mass of the star.

\begin{figure}
\begin{center}
\includegraphics[scale=0.5]{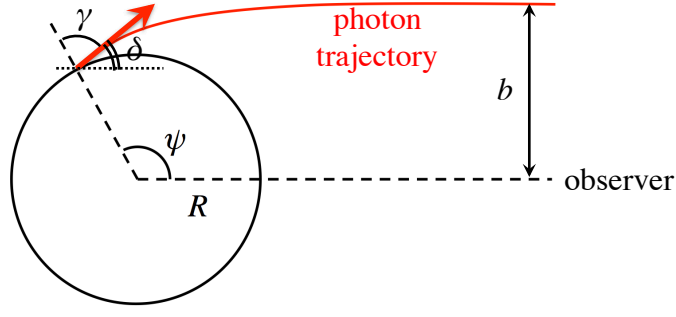} 
\end{center}
\caption{
Image for the photon trajectory radiated from the stellar surface.
}
\label{fig:trajectory}
\end{figure}

\begin{figure}
\begin{center}
\includegraphics[scale=0.5]{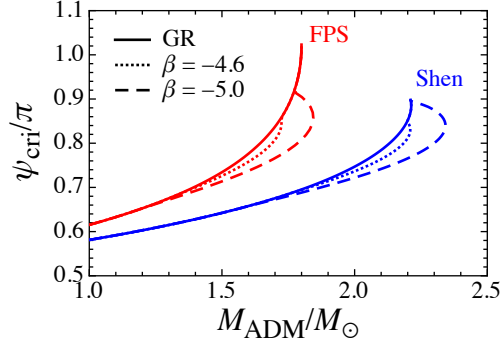} 
\end{center}
\caption{
Critical values of $\psi$ when $\gamma$ becomes equivalent to $\pi/2$ are shown as a function of the ADM mass for FPS and Shen EOSs. The solid line denotes the values in general relativity, while the dotted and dashed lines correspond to those in scalar-tensor gravity with $\beta=-4.6$ and $-5.0$.
}
\label{fig:psic}
\end{figure}

The photon trajectory radiated from the neutron star surface is determined from the Euler-Lagrange equation on $\theta=0$ plane. In fact, considering the situation shown in Fig. \ref{fig:trajectory}, the angle $\psi$ for a specific stellar model is determined via 
\begin{equation}
  \psi(R) = \int_R^\infty \frac{dr}{\tilde{C}}\left[\frac{1}{\tilde{A}\tilde{B}}\left(\frac{1}{b^2}-\frac{\tilde{A}}{\tilde{C}}\right)\right]^{-1/2},  \label{eq:psi}
\end{equation}
where $b$ is the impact parameter given with the emission angle $\gamma$ as 
\begin{equation}
  \sin\gamma = b\sqrt{\frac{\tilde{A}(R)}{\tilde{C}(R)}}. \label{eq:impact}
\end{equation}
The bending angle $\delta$ is defined with the angles $\psi$ and $\gamma$ as $\delta=\psi-\gamma$. Combining Eq. (\ref{eq:psi}) with Eq. (\ref{eq:impact}), one can numerically obtain the relation between $\psi$ and $\gamma$ \cite{SM2017}. The emission angle $\gamma$ increases with $\psi$, but the maximum value of $\gamma$ should be $\pi/2$ in order for the photon from the stellar surface to be observed. The angle $\psi$ when $\gamma=\pi/2$ is particularly expressed as $\psi_{\rm cri}$. In Fig. \ref{fig:psic}, we show the value of $\psi_{\rm cri}$ as a function of the ADM mass of the stellar models constructed with FPS and Shen EOSs, where the solid line denotes the values in general relativity, while the dotted and dashed lines correspond to those in scalar-tensor gravity with $\beta=-4.6$ and $-5.0$. Focusing on the stellar models  with $M_{\rm ADM}=1.6M_\odot$, the value of $\psi_{\rm cri}$ for FPS EOS in scalar-tensor gravity becomes $2.3\%$ with $\beta=-4.6$, $3.8\%$ with $\beta=-4.8$, and $5.0\%$ with $\beta=-5.0$ smaller than that in general relativity, while $\psi_{\rm cri}$ for Shen EOS in scalar-tensor gravity with $\beta=-5.0$ becomes $0.04\%$ smaller than that in general relativity. We remark that the stellar models with $M_{\rm ADM}=1.6M_{\odot}$ for Shen EOS in scalar-tensor gravity with $\beta=-4.6$ and $-4.8$ are equivalent to that in general relativity. On the other hand, for the stellar models with $M_{\rm ADM}=2.1M_{\odot}$ constructed with Shen EOS, $\psi_{\rm cri}$ becomes $2.6\%$ with $\beta=-4.6$, $4.2\%$ with $\beta=-4.8$, and $5.4\%$ with $\beta=-5.0$ smaller than that in general relativity. This is mainly because of the reduction of stellar compactness shown in the right panel of Fig. \ref{fig:MR} due to the scalarization. 


\section{Pulse profile from neutron stars}
\label{sec:IV}

Now, assuming the existence of the hot spot on the neutron star, we consider the situation that the photon radiates from the surface element $dS$ with the surface radiation intensity $I_0(\gamma)$, where the distance between the neutron star and the observer is $D$ in $r$ coordinate. We remark that the physical distance between the neutron star and the observer, $\tilde{D}$, should be determined via the relationship of $(\tilde{R}+\tilde{D})^2=\tilde{C}(R+D)$, i.e., $\tilde{D}=(R+D)\exp[\beta\varphi_{\rm ob}^2/2]-\tilde{R}$ with $\varphi_{\rm ob}\equiv \varphi(R+D)$. Then, the observed flux $d\tilde{F}$ is calculated by
\begin{equation}
  d\tilde{F} = I_0(\gamma) \tilde{A}(R)\tilde{C}(R)\cos\gamma\frac{dS}{R^2 D^2} \frac{d(\cos\gamma)}{d\mu}, \label{eq:dF}
\end{equation}
where $\mu=\cos\psi$ \cite{SM2017}. Adopting a pointlike spot approximation for simplicity \cite{Beloborodov2002} and assuming the black body emission from the hot spot with the isotropic intensity expressed by $I_0(\gamma)=I_0$, the observed flux $\tilde{F}$ from the hot spot is obtained by integrating Eq. (\ref{eq:dF}), i.e.,
\begin{equation}
   \tilde{F} = F_1\frac{\sin\gamma\,\cos\gamma}{\sin\psi}\frac{d\gamma}{d\psi},   \label{eq:FF}
\end{equation}
where $F_1\equiv I_0 s\tilde{A}(R)\tilde{C}(R)/(R^2D^2)$ and $s$ denotes the spot area given by $s=\int dS$. In particular, we consider a neutron star with two (antipodal) hot spots associated with the polar caps of the stellar magnetic field, where the hot spot closer to the observer is referred to as ``primary" and the other spot is ``antipodal".

\begin{figure*}
\begin{center}
\begin{tabular}{cc}
\includegraphics[scale=0.5]{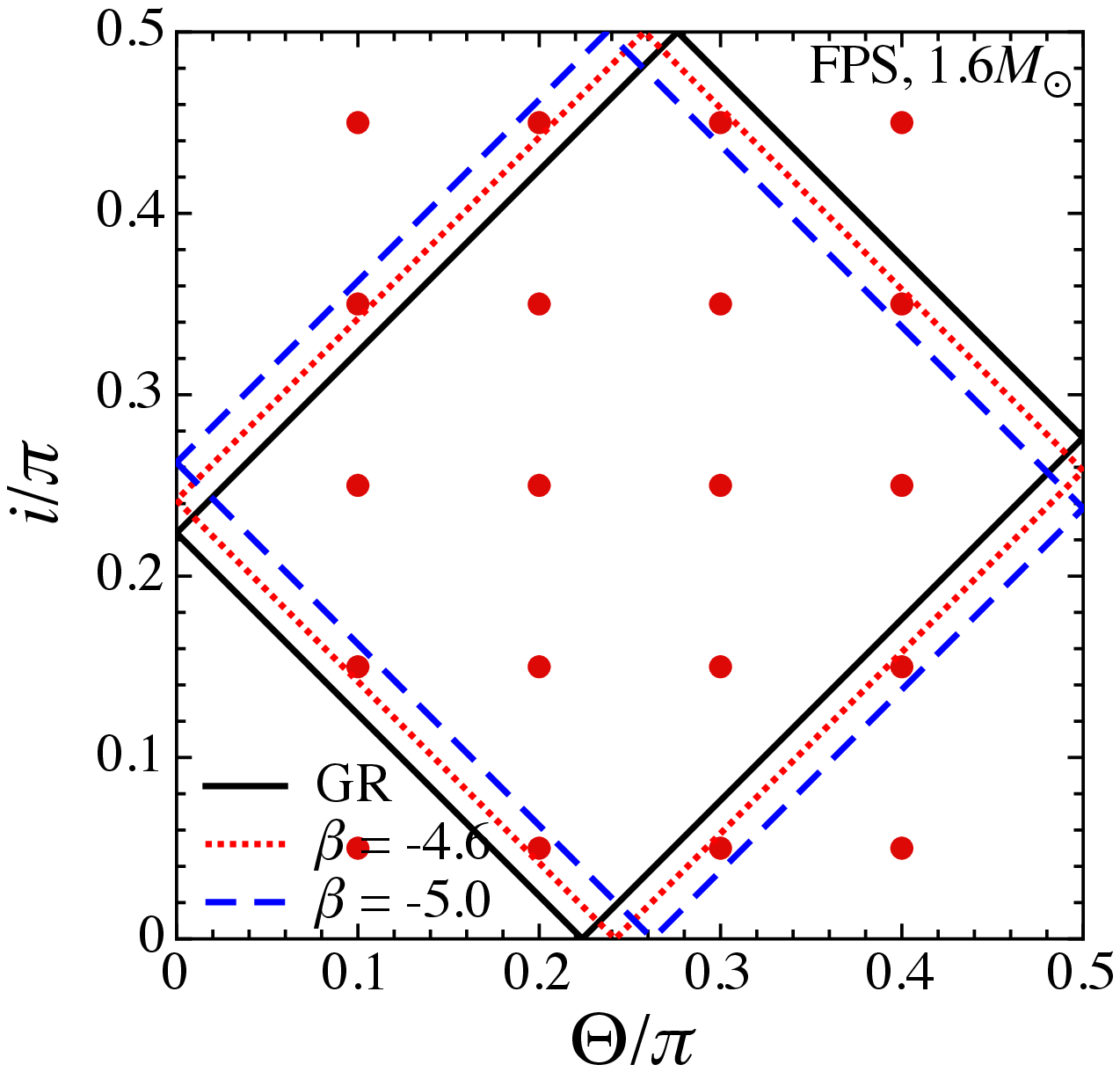} &
\includegraphics[scale=0.5]{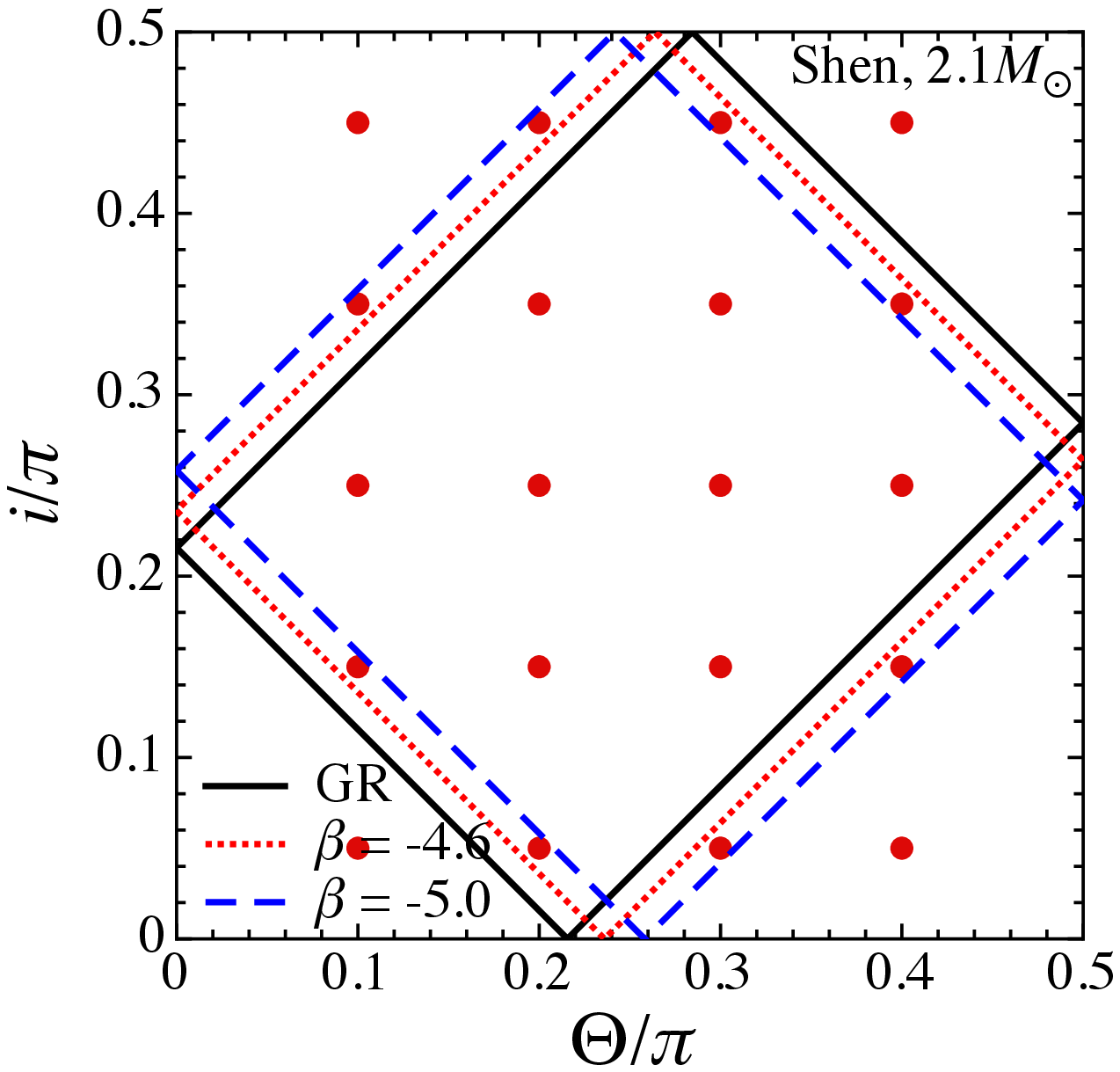} 
\end{tabular}
\end{center}
\caption{
The boundaries, which classify the observations of two hot spots, are shown with the solid line in general relativity, and with the dotted and dashed lines in scalar-tensor gravity with $\beta=-4.6$ and $-5.0$. The left and right panels correspond to the stellar models with $1.6M_\odot$ for FPS EOS and with $2.1M_\odot$ for Shen EOS, respectively. For each case, the boundary is determined by connecting $(\Theta/\pi,i/\pi)=(0,1-\psi_{\rm cri}/\pi)$, $(1-\psi_{\rm cri}/\pi,0)$, $(0.5,\psi_{\rm cri}/\pi-0.5)$, and $(\psi_{\rm cri}/\pi-0.5,0.5)$ with $\psi_{\rm cri}$ determined for each stellar model. The filled circles in the figure denote the specific values of $(\Theta/\pi,i/\pi)$, which are adopted for considering the pulse profiles.
}
\label{fig:i-theta}
\end{figure*}

For considering the observation of the pulse profile from the rotating neutron star with the angular velocity $\Omega$, one has to introduce two specific angles, i.e., $\Theta$, which is the angle between the rotational and the magnetic axes, and $i$, which is the angle between the rotational axis and the direction to the observer, where $\Theta$ and $i$ can be chosen in such a way that $i \le\pi/2$ and $\Theta\le\pi/2$ (see Fig. 3 in Ref. \cite{SM2017}). Then, the value of $\mu=\cos\psi$ for the primary hot spot is varying with time as
\begin{equation}
  \mu(t) = \sin i\sin\Theta\cos(\Omega t) + \cos i\cos\Theta, \label{eq:mu}
\end{equation}
because the value of $\mu$ is determined via $\mu=\bm{n}_{\rm p} \cdot \bm{d}$ with the normal vector at the primary hot spot ($\bm{n}_{\rm p}$) and the unit vector pointing toward the observer ($\bm{d}$) \cite{SM2017,Beloborodov2002}, where we put $t=0$ when the primary hot spot comes to the closest point to the observer, i.e., $\psi$ becomes minimum. With respect to the antipodal hot spot, due to the symmetry of the system, the normal vector at the antipodal hot spot ($\bm{n}_{\rm a}$) is given by $\bm{n}_{\rm a}=-\bm{n}_{\rm p}$. Thus, the value of $\mu$ for the antipodal hot spot ($\bar{\mu}$) is determined by $\bar{\mu}(t)=\bm{n}_{\rm a}\cdot \bm{d} = -\mu(t)$ as a function of time. Here, for simplicity we assume that $\Omega$ is small enough to neglect the frame dragging effect due to the stellar rotation. In fact, the frame dragging effect could be neglected for $1/T\lesssim $ a few hundred Hz \cite{PO2014}, where $T$ denotes the rotational period, i.e., $T=2\pi/\Omega$. Then, the pulse profile with the given angles $\Theta$ and $i$ becomes periodic, where the amplitude at $t/T$ for $0.5\le t/T\le 1$ becomes the same as that at $1-t/T$. So, the figures for the pulse profile from each stellar model as shown below, are plotted in the period of $0\le t/T\le 0.5$. In addition, from Eq. (\ref{eq:mu}), one can see that the pulse profile with $(\Theta,i)=(\theta_1,\theta_2)$ is equivalent to that with $(\Theta,i)=(\theta_2,\theta_1)$ for $0\le\theta_i\le\pi/2$ with $i=1$ and 2.

The primary (antipodal) hot spot can be seen when $\mu$ $(\bar{\mu})$ is larger than $\cos\psi_{\rm cri}$. Thus, the flux from the primary (antipodal) hot spot, $\tilde{F}$ ($\tilde{F}_{\rm a}$), is calculated via Eq. (\ref{eq:FF}) for $\mu$ $(\bar{\mu})$ $\ge \cos\psi_{\rm cri}$ and becomes zero for $\mu$ $(\bar{\mu})$ $\le \cos\psi_{\rm cri}$. With $\tilde{F}(t)$ and $\tilde{F}_{\rm a}(t)$, the observed flux from the pulsar is determined as
\begin{equation}
  F_{\rm ob}(t) = \tilde{F}(t) + \tilde{F}_{\rm a}(t).
\end{equation}
Depending on the angles $\Theta$ and $i$, the pulse shape from the pulsar can be classified into four cases, where the boundary between different cases is determined by $\psi_{\rm cri}$ \cite{SM2017,Beloborodov2002}. That is, in the $\Theta/\pi-i/\pi$ plane for $0\le i/\pi\le 0.5$ and $0\le \Theta/\pi\le 0.5$, (i) only the primary spot can be observed at any time, where $i + \Theta < \pi -\psi_{\rm cri}$, (ii) the primary spot can be observed at any time and the antipodal spot can be also observed sometimes, where $\pi -\psi_{\rm cri} < i + \Theta < \psi_{\rm cri}$ and $\psi_{\rm cri}-\pi<i-\Theta<\pi-\psi_{\rm cri}$, (iii) only primary spot can be observed, or the both spots can be observed, or only the antipodal spot can be observed, where $i + \Theta > \psi_{\rm cri}$, and (iv) the both spots can be observed at any time, where $i-\Theta<\psi_{\rm cri}-\pi$ or $i-\Theta>\pi-\psi_{\rm cri}$. In Fig. \ref{fig:i-theta}, we show such a classification for the stellar models constructed with FPS EOS with $1.6M_\odot$ in the left panel and for those constructed with Shen EOS with $2.1M_\odot$ in the right panel, where the solid line denotes the result in general relativity while the dotted and dashed lines denote those in scalar-tensor gravity with $\beta=-4.6$ and $-5.0$. Since $\psi_{\rm cri}$ depends on the gravitational theory even for fixing the stellar mass (as shown in Fig. \ref{fig:psic}), the boundary of the classification in the $\Theta/\pi-i/\pi$ plane also depends on the gravitational theory.

\begin{figure*}
\begin{center}
\begin{tabular}{c}
\includegraphics[scale=0.38]{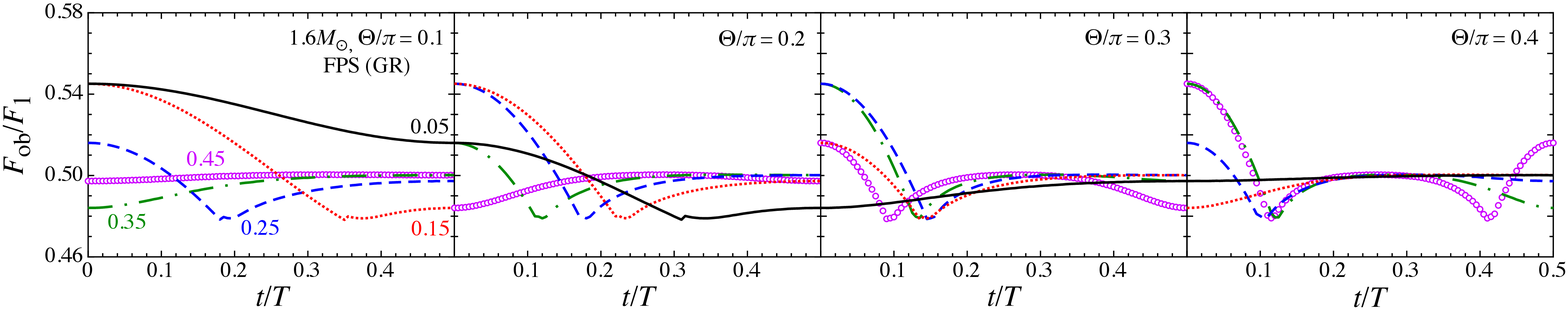} \\
\includegraphics[scale=0.38]{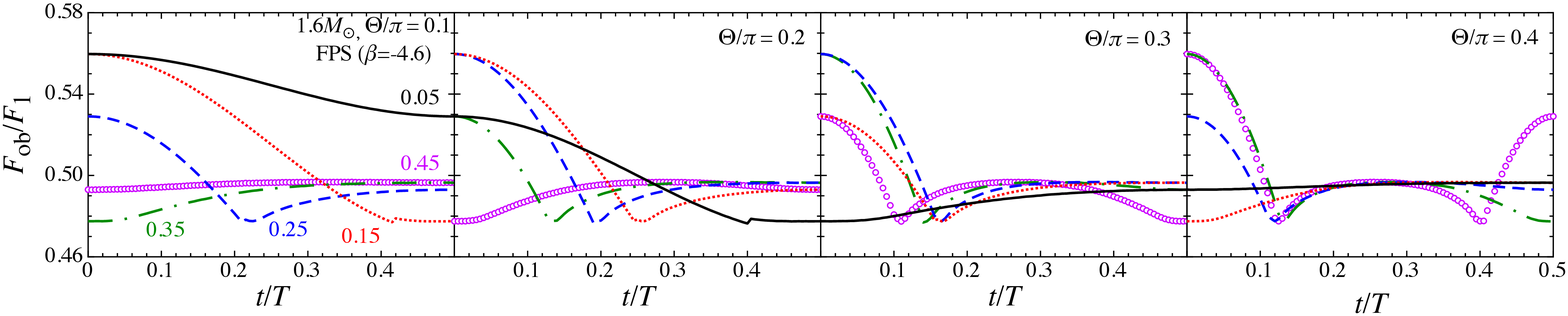} \\
\includegraphics[scale=0.38]{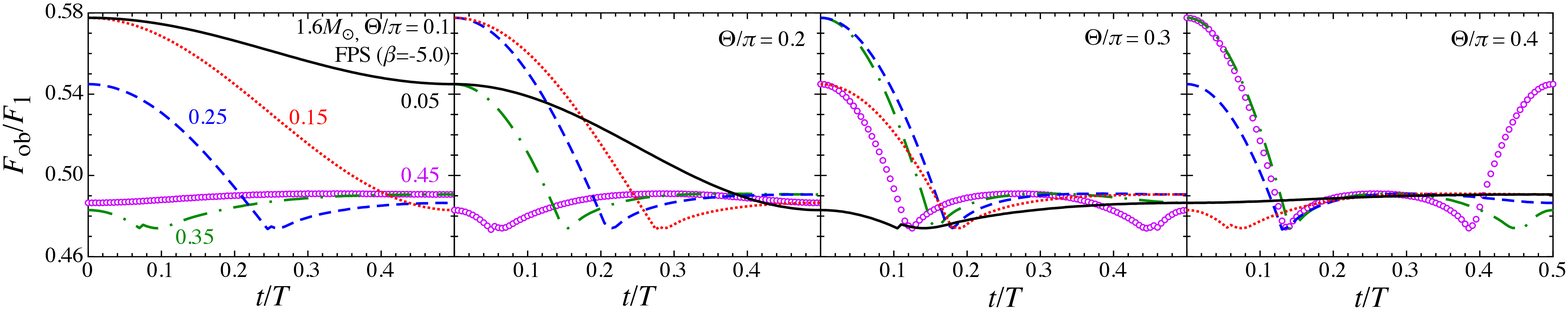} 
\end{tabular}
\end{center}
\caption{
Pulse profiles from the stellar models for FPS EOS with $M_{\rm ADM}=1.6M_\odot$ in general relativity, i.e., $\beta=0$, (upper row), in scalar-tensor gravity with $\beta=-4.6$ (middle row), and with $\beta=-5.0$ (lower row). For each stellar model, the panels from left to right are the results for $\Theta/\pi=0.1$, $0.2$, $0.3$, and $0.4$. In each panels, the solid, dotted, dashed, dot-dashed lines, and open circles respectively denote the results for $i/\pi=0.05$, $0.15$, $0.25$, $0.35$, and $0.45$.
}
\label{fig:pulse-FPSM16}
\end{figure*}

\begin{figure*}
\begin{center}
\includegraphics[scale=0.38]{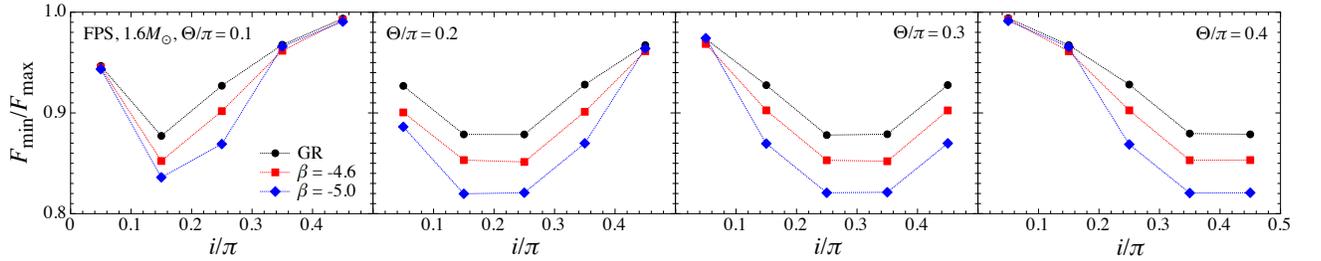}
\end{center}
\caption{
The ratio of the minimum observed flux to maximum observed flux shown in Fig. \ref{fig:pulse-FPSM16} is shown for various angles of $i/\pi$ and $\Theta/\pi$. In each panel, the circles denote the results in general relativity, while the squares and diamonds denote the results in scalar-tensor gravity with $\beta=-4.6$ and $-5.0$, respectively.
}
\label{fig:FF1}
\end{figure*}

In Fig. \ref{fig:pulse-FPSM16}, we plot the pulse profiles from the neutron star for FPS EOS with $M_{\rm ADM}=1.6M_\odot$ in general relativity (upper row), in scalar-tensor gravity with $\beta=-4.6$ (middle row), and with $\beta=-5.0$ (lower row) for various values of $(\Theta/\pi, i/\pi)$, which correspond to the dots in the left panel of Fig. \ref{fig:i-theta}. One may observe that the pulse shapes are relatively similar to each other with the same angles of $i/\pi$ and $\Theta/\pi$. So, with the direct observation of the pulse shape one may make a halfdecent constraint on the combination of the angles $i/\pi$ and $\Theta/\pi$ independent from the gravitational theory. On the other hand, one can also see that the ratio of the minimum amplitude to the maximum amplitude ($F_{\rm min}/F_{\rm max}$) in the pulse profile for given angles of $i/\pi$ and $\Theta/\pi$ depends on the coupling constant $\beta$. To clarify this point, we show such a ratio in Fig. \ref{fig:FF1} for the pulse profiles shown in Fig. \ref{fig:pulse-FPSM16}. Here, the circles correspond to the ratio expected in general relativity, while the squares and diamonds correspond to that in scalar-tensor gravity with $\beta=-4.6$ and $-5.0$. From this figure, one can see that the cases with $(\Theta/\pi,i/\pi)=(0.1,0.05)$, $(0.1,0.35)$, $(0.1,0.45)$, $(0.2,0.45)$, $(0.3,0.05)$, $(0.4,0.05)$, and $(0.4,0.15)$ are almost independent from the gravitational theory, while the other cases depend strongly on the coupling constant $\beta$. Comparing this result with the left panel in Fig. \ref{fig:i-theta}, the angles with which $F_{\rm min}/F_{\rm max}$ depends strongly on $\beta$, are in the region where $i + \Theta > \pi -\psi_{\rm cri}$ and $\psi_{\rm cri}-\pi<i-\Theta<\pi-\psi_{\rm cri}$. That is, one can see the dependence of the coupling constant $\beta$ in $F_{\rm min}/F_{\rm max}$ for the combination of the angles of $\Theta/\pi$ and $i/\pi$ when the antipodal hot spot can be sometimes (not always) observed, i.e., the situation (ii) or (iii) explained the above.

\begin{figure*}
\begin{center}
\begin{tabular}{c}
\includegraphics[scale=0.38]{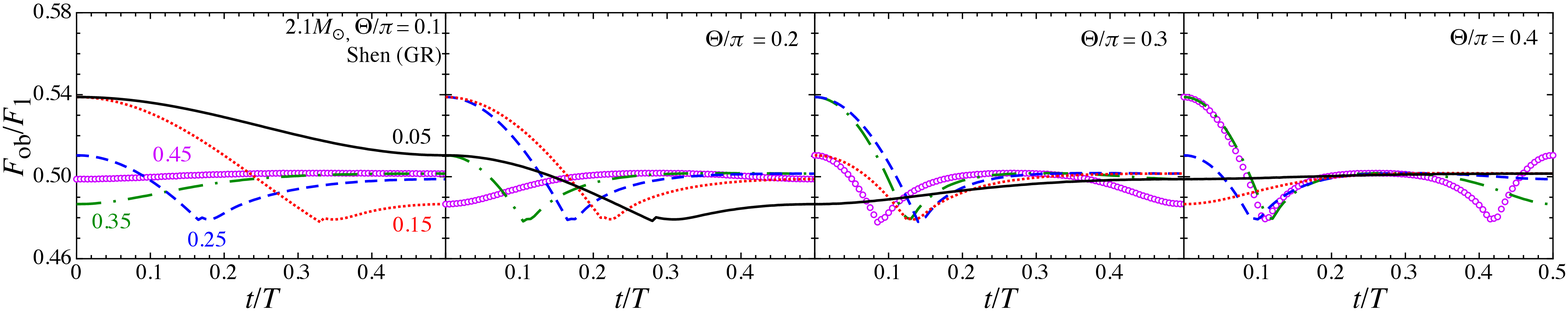} \\
\includegraphics[scale=0.38]{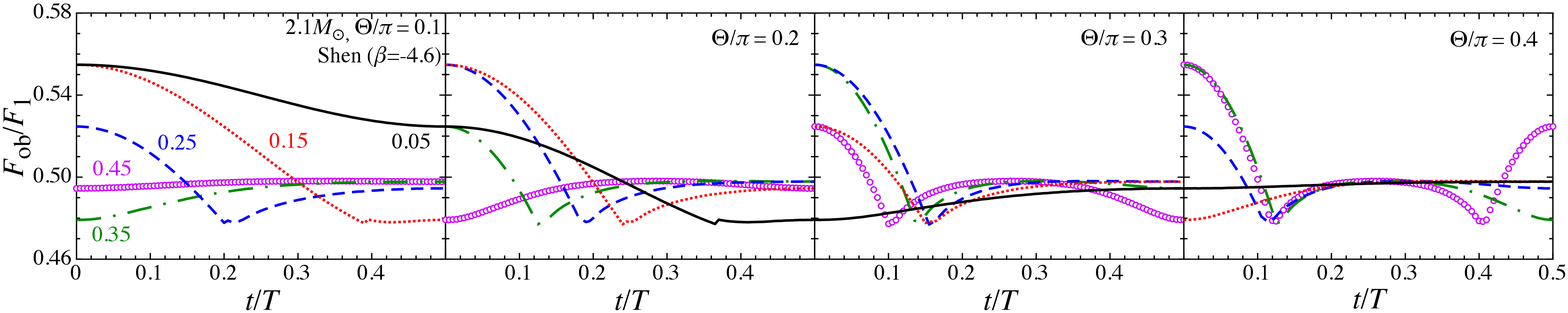} \\
\includegraphics[scale=0.38]{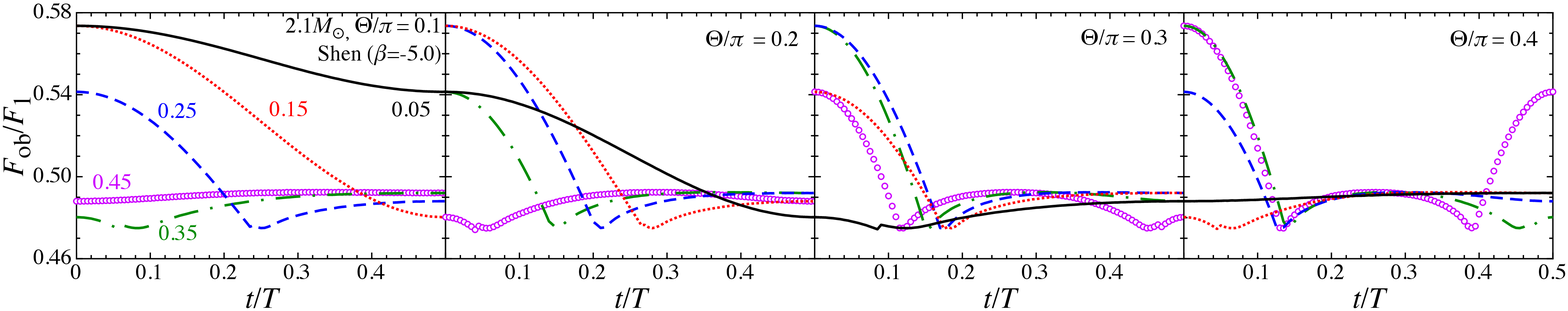} 
\end{tabular}
\end{center}
\caption{
Same as Fig. \ref{fig:pulse-FPSM16}, but for Shen EOS with $M_{\rm ADM}=2.1M_\odot$.
}
\label{fig:pulse-ShenM21}
\end{figure*}

\begin{figure*}
\begin{center}
\includegraphics[scale=0.38]{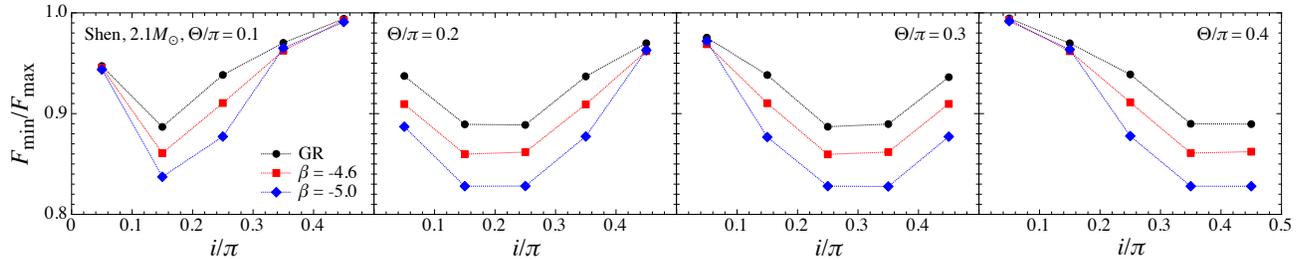} 
\end{center}
\caption{
The ratio of the minimum observed flux to maximum observed flux shown in Fig. \ref{fig:pulse-ShenM21} is shown for various angles of $i/\pi$ and $\Theta/\pi$. In each panel, the circles denote the results in general relativity, while the squares and diamonds denote the results in scalar-tensor gravity with $\beta=-4.6$ and $-5.0$, respectively.
}
\label{fig:FF2}
\end{figure*}

In the same way, in Fig. \ref{fig:pulse-ShenM21} we show the pulse profiles from the neutron star for Shen EOS with $M_{\rm ADM}=2.1M_\odot$ in general relativity (upper row), in scalar-tensor gravity with $\beta=-4.6$ (middle row), and with $\beta=-5.0$ (lower row) for various values of $i/\pi$ and $\Theta/\pi$, which correspond to the dots in the right panel of Fig. \ref{fig:i-theta}. In addition, the resultant ratio $F_{\rm min}/F_{\rm max}$ is shown in Fig. \ref{fig:FF2}. Again, we find that the pulse shapes expected in general relativity may be more or less similar to those from the neutron star with the same mass expected in scalar-tensor gravity, for the given angles $i/\pi$ and $\Theta/\pi$. On the other hand, the ratio of the minimum amplitude to maximum amplitude in the pulse profile depends strongly on the coupling constant $\beta$ when the antipodal hot spot can be sometimes observed, while such a ratio is almost independent from $\beta$ either when only the primary spot can be observed or when the both spots can be observed in any time. We remark that the results shown in Figs. \ref{fig:pulse-FPSM16} and \ref{fig:FF1} are very similar to those shown in Figs. \ref{fig:pulse-ShenM21} and \ref{fig:FF2}, because the pulse profile (or the light bending) depends on the stellar compactness and the stellar compactness for the neutron star constructed with FPS EOS with $1.6M_\odot$ is very similar to that with Shen EOS with $2.1M_\odot$ (see the right panel in Fig. \ref{fig:MR}). Thus, via the simultaneous observations of the pulse profile and the stellar mass, one may constrain the EOS for neutron star matter.

\begin{figure*}
\begin{center}
\begin{tabular}{c}
\includegraphics[scale=0.38]{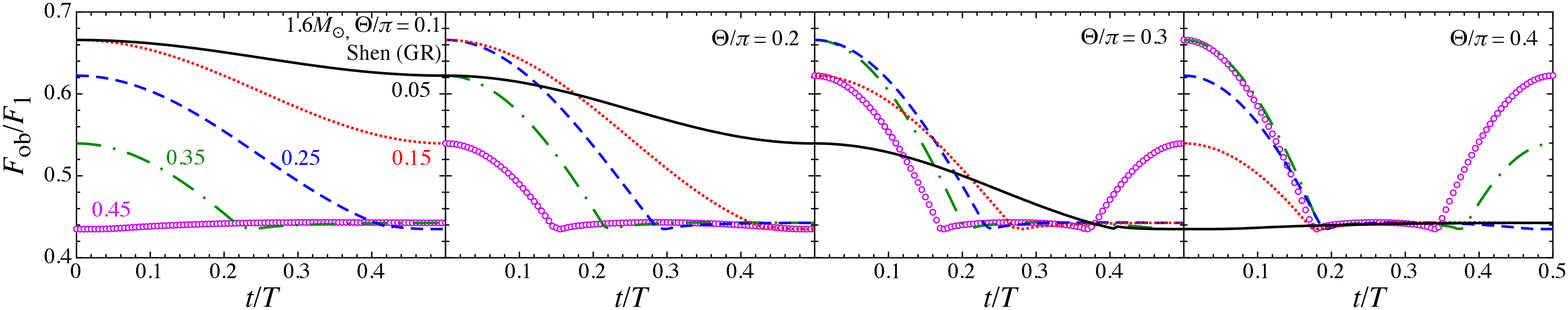} \\
\includegraphics[scale=0.38]{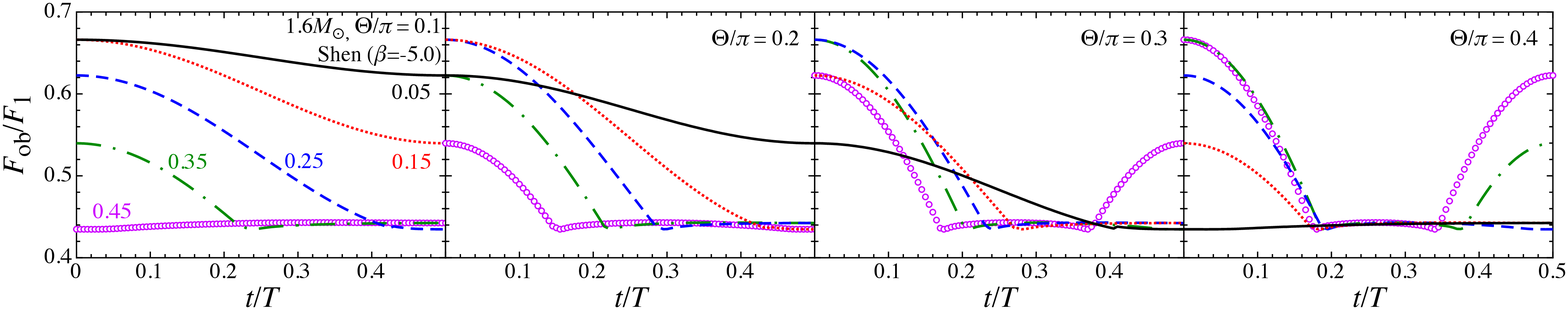} 
\end{tabular}
\end{center}
\caption{
Same as Fig. \ref{fig:pulse-FPSM16}, but for Shen EOS with $M_{\rm ADM}=1.6M_\odot$. We remark that since the stellar models in scalar-tensor gravity with $\beta=-4.6$ are the same as in general relativity, we omit such stellar models. 
}
\label{fig:pulse-ShenM16}
\end{figure*}

\begin{figure*}
\begin{center}
\includegraphics[scale=0.38]{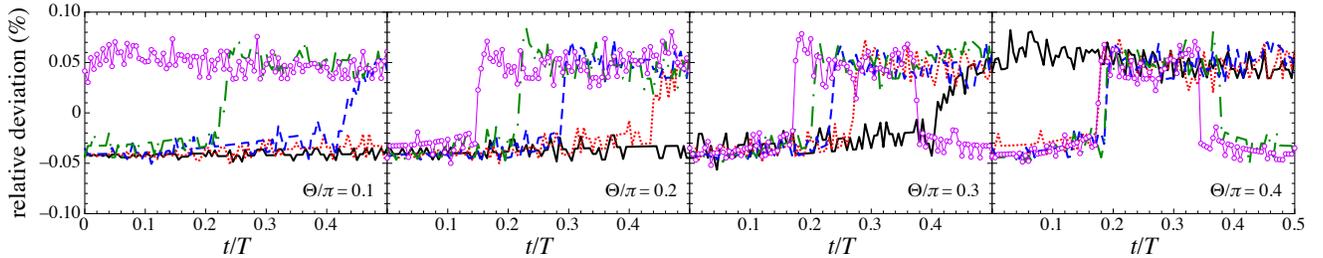} 
\end{center}
\caption{
Relative deviation of the amplitude of pulse profile expected in scalar-tensor gravity with $\beta=-5.0$ from those in general relativity for the stellar model constructed with Shen EOS and $M_{\rm ADM}=1.6M_\odot$ for the various angles $i/\pi$ and $\Theta/\pi$, i.e., $(F_{\rm GR} - F_{\rm ST})/F_{\rm GR}\times 100$ with $F_{\rm GR}\equiv F_{\rm ob}/F_1$ in general relativity and  $F_{\rm ST}\equiv F_{\rm ob}/F_1$ in scalar-tensor gravity with $\beta=-5.0$, which are shown in Fig. \ref{fig:pulse-ShenM16}. The meaning of the various lines is the same as in Fig. \ref{fig:pulse-ShenM16}.
}
\label{fig:relative}
\end{figure*}

Furthermore, in Fig. \ref{fig:pulse-ShenM16} we show the pulse profiles from the neutron star for Shen EOS with $M_{\rm ADM}=1.6M_\odot$ in general relativity (upper row) and in scalar-tensor gravity with $\beta=-5.0$ (lower row). As shown in Fig. \ref{fig:MR}, the neutron star in scalar-tensor gravity with $\beta=-5.0$ is scalarized, i.e., the non-zero scalar field exists inside/around the neutron star, even though the stellar radius is almost the same as that for the stellar model with the same mass expected in general relativity. Even so, as shown in Fig. \ref{fig:pulse-ShenM16}, the pulse profiles in general relativity are the same as those in scalar-tensor gravity. Thus, we find that the pulse profile is almost independent from the existence of the scalar field, while the stellar compactness is crucial for determining the profile. We remark that one can see the effect of the existence of a scalar field in the pulse profile as shown in Figs. \ref{fig:FF1} and \ref{fig:FF2}, but such an effect can be seen as a result of the difference of the stellar radius for the given mass due to the existence of the scalar field. Additionally, in Fig. \ref{fig:relative} we show the relative deviation of the amplitude of pulse profile expected in scalar-tensor gravity with $\beta=-5.0$ from those in general relativity for the stellar model constructed with Shen EOS and $M_{\rm ADM}=1.6M_\odot$ for the various angles $i/\pi$ and $\Theta/\pi$, where the meaning of the different lines is the same as in Fig. \ref{fig:pulse-ShenM16}. Depending on the angles $i/\pi$ and $\Theta/\pi$, the relative deviation can vary a little with time, but it is only less than $0.1$ \%.

\section{Conclusion}
\label{sec:V}

We examine the pulse profiles from a neutron star in scalar-tensor gravity, assuming the antipodal hot spots based on the polar cap model. We find that the pulse profile depends strongly on the stellar compactness, where the existence of the scalar field does not directly change the pulse shape. That is, the pulse shape from the neutron star without the scalar field (or the neutron star in general relativity) is almost the same as that with the scalar field (or the scalarized neutron star in scalar-tensor gravity) if the both stellar models have similar compactness. The pulse profile also depends on the EOS for neutron star matter, while the pulse profile with a specific EOS is very similar to that with another EOS if the stellar compactness is similar to each other. So, the observation of the pulse profile with the help of the additional observation of the stellar mass could tell us the EOS. With a specific EOS, the pulse shape from the neutron star in general relativity is more or less similar to that in scalar-tensor gravity for the given angles $\Theta$ and $i$, where $\Theta$ is the angle between the rotational and magnetic axes and $i$ is the angle between the rotational axis and the direction to the observer. On the other hand, the ratio of the minimum amplitude to the maximum amplitude in the pulse profile depends on the coupling constant in the gravitational theory, only if the antipodal hot spot can be seen not always but sometimes. This dependence on the coupling constant might enable us to observationally reveal the gravitational theory in strong field regime.

In this article, we completely omit the effect of the pulsar magnetosphere on the photon propagation. The emission from the pulsar magnetosphere must be non-thermal, which strongly depends on the magnetic field structure, and may be observed as a background. That is, the energy spectrum for the thermal emission of the pulse profiles is different from that of the background non-thermal emission. In any way, assuming a theoretical model of magnetosphere or via another way, one would have to remove the effect of non-thermal emission form the observed spectra.  Additionally, one may not observe the periodicity in the emission from the magnetosphere, if such emission comes from not only the hot spots but also the other parts around the surface. So, since the signal from the hot spots we considered in this paper are independent from the frequencies in X-ray observations, one might be able to distinguish the signal from the hot spots from the noise from the magnetosphere by checking the periodicity in the X-ray signals as changing the frequencies.

\acknowledgments
This work was performed in part at Aspen Center for Physics, which is supported by National Science Foundation grant PHY-1607611. This work was also supported in part by Grant-in-Aid for Scientific Research (C) through Grant No. 17K05458 provided by JSPS.



\end{document}